# Methodology of safety evaluation of In-Vessel Retention

F. Fichot[1], L. Carénini[1], S. Brumm[2], M. Sangiorgi[3]
[1]: IRSN, BP3 – 13115 St Paul lez Durance, France
[2]: EC/JRC, Petten, The Netherlands
[3]: EC/JRC, Ispra, Italy

florian.fichot@irsn.fr, laure.carenini@irsn.fr, stephan.brumm@ec.europa.eu, marco.sangiorgi@ec.europa.eu

## 1  Introduction

Molten corium stabilization following a severe accident is of crucial importance in order to ensure containment integrity on a long-term basis and minimizing radioactive elements releases outside the plant. Among the possible options, In-Vessel Retention (IVR) through external cooling appears as an attractive solution that would limit the dispersion of corium in the plant and minimize the risks of containment failure. Nevertheless its feasibility has to be proved.

The IVR strategy is already adopted in Europe for some VVER 440 type 213 reactors thanks to thorough research work started in the '90s for the Finnish Loviisa power plant, and subsequently extended to Bohunice and Mochovce (Slovakia), Dukovany (Czech Republic) and Paks (Hungary) power plants. The strategy is also included in the design of some high power new Gen.III reactors such as AP1000, APR 1400 and Chinese HPR1000 and CAP1400. It has also been studied in the past for other reactor concepts like KERENA (1250 MWe - BWR), AP600 or VVER-640.

Current approaches for reactors with relatively small power, such as VVER 440 or AP600, use conservative assumptions for the safety demonstration. However, for higher power reactors (around 1000 MWe), the safety margin is reduced and it is necessary to evaluate the IVR strategy with best-estimate methods in order to reduce the uncertainties associated with the involved phenomena. Additional R&D as well as a revision of the methodology are needed to ensure and demonstrate adequate safety margins, including, in particular, best-estimate evaluations of thermal load applied on the vessel and mechanical resistance of the ablated vessel.

The IVMR project (In-Vessel Melt Retention) was built with the goal of providing new knowledge (experimental, theoretical and technical) and a new methodology able to provide a best-estimate evaluation of IVR strategy for large power reactors. The main objective of Task 2.1 within WP2 was to define a common methodology to analyse IVR Severe Accident Management (SAM) strategy for the different types of EU NPPs. It started by reviewing the status of existing methodology and aimed at elaborating a more general, updated and less conservative one applicable to several types of reactors.

This paper describes the proposed new methodology. It starts with the identification of the deficiencies of the standard methodology when it is applied to a high power reactor. It introduces the minimum vessel thickness as a parameter representing the cumulated imbalance between internal heat load and external cooling. Then it explains how to use that parameter in the evaluation of the safety margin. Although some examples are given as illustrations, it must be kept in mind that this paper proposes a generic methodology but there cannot be any generic conclusion: any reactor design must be evaluated independently.

## 2  The possible roles of IVR and ERVC: mitigation or termination of the accident progression

The terminology "In-Vessel Retention" or IVR was initially proposed to describe a strategy that would lead to the termination of corium progression by using the lower part of the vessel as a heat exchanger through which it would be possible to extract the residual power and progressively stabilize the corium. For the investigated cases (Loviisa plant, for instance (Kymäläinen et al., 1997)), it was proved that the vessel would not fail, because the remaining vessel thickness was estimated to be most likely around 10 cm and, in any case, always larger than 6 cm. With this statement established, the remaining elements of the safety demonstration are essentially related to the design (reliability and performance of the implemented hardware): depressurization system, efficient external cooling system, efficient containment condensation system, etc.

When investigating the possible IVR application to higher power reactors (such as AP1000 (Esmaili and Khatib-Rahbar, 2005) or APR1400 (Rempe et al., 2004)), a new point had to be considered in the safety analysis: the probability of vessel failure was not zero and not even negligible. This led to shifting the safety assessment to the evaluation of the risk of containment failure due to the energetic interaction of corium with water after vessel failure. In that case, the main issue was not "Can the corium be retained within the vessel?" but "Can the containment withstand a possible steam explosion in case of vessel failure?" If the answer to this question is positive, then it is acceptable to flood the external part of the vessel in order to either stop the corium progression in the vessel or delay the vessel failure. Another consequence, in that case, is that additional measures must be implemented to ensure the stabilization of corium outside the vessel (core catcher or corium spreading and/or sacrificial concrete for example).

Therefore, we see that there are two different uses of External Reactor Vessel Cooling (ERVC):

1. ERVC to **prevent** vessel failure: in that case, it is right to talk about in-vessel retention of corium and the objective of safety evaluation is to evaluate the risk of vessel failure and show that it is practically eliminated (i.e. extremely unlikely with a high degree of confidence).

2. ERVC to **prevent or delay** vessel failure: in that case, the objective is not in-vessel retention but it is a defence-in-depth approach where the possibility to stop the corium progression inside the vessel is considered with an appropriate system but it is not the only possibility and ex-vessel systems are also implemented. The objective of safety evaluation is to identify the cases leading to vessel failure, to evaluate the associated risk of containment failure, and show that it is negligible. If the probability of vessel failure is too high, it may even lead to the conclusion that implementation of ERVC does not provide a significant improvement of the safety of the reactor and is not necessary.

In the worst cases, depending on the reactor design, the implementation of ERVC may even increase the risk of containment damage or failure (Fichot et al., 2014).

So, it must be kept in mind that ERVC implementation is not equivalent to IVR and that it may be only a safety measure in a chain of other measures, following the principle of defence-in-depth. The association of the two terminologies (ERVC and IVR), which is frequently done, may be confusing or misleading.

Therefore, one of the key issues in the safety analysis of a reactor where ERVC is implemented is to determine if the probability of vessel failure is extremely unlikely with a high degree of confidence, or if it is not negligible. **The two following cases correspond to two different approaches of using ERVC mentioned above**. In the first case, the probability of vessel failure must be evaluated with the highest possible accuracy. In the second case, the evaluation does not need to be very accurate. In-between those two cases an arbitrary threshold of residual risk can be chosen to accept the IVR strategy or to require additional measures for ex-vessel stabilization of corium. This threshold should be examined on a case-by-case basis, depending on the reactor design.

The following table summarizes the possible situations:

| Probability of vessel failure | Safety assessment conclusion | Accuracy needed |
|---|---|---|
| **Negligible** | IVR can be assumed, with a residual risk | High |
| **Not negligible** | IVR is not guaranteed. Ex-vessel stabilization must be implemented too. | Medium |

## 3    Background

The methodology introduced initially for the AP600 (Rempe et al., 1997; Theofanous et al., 1997a) and the Loviisa plant (Kymäläinen et al., 1997; Kymäläinen and Tuomisto, 2000) included a probabilistic evaluation of the maximum heat flux and minimum thickness of the vessel. The analysis was made with a single configuration of corium and metal in the lower plenum: the entire core was assumed to be molten, forming the oxide pool and all the internal steel structures and ablated vessel were assumed to be included in the top metal layer, above the oxide pool. This configuration was initially called "bounding situation" by (Theofanous et al. 1997). In their paper, they explain that "this configuration is of fundamental significance in assessing the problem at hand, not only because it bounds all intermediate states, but also because it represents the final state that would actually be realized in any in-vessel retention scenario". But this statement is proven wrong by today's phenomenological modelling. Most of the reactor calculations performed by IVMR partners during the project have shown that:

•    The bounding case does not bound all intermediate states: there are transient situations where the peak heat flux is higher than the heat flux at final state.

•    The bounding case does not represent the final state: the metal and oxide layers masses may vary, depending on assumptions about densities and thermochemical equilibrium compositions.

Moreover, the shape of the ablated vessel obtained after a transient may significantly differ from the shape deduced from the final state.

These statements are illustrated in Figure 1 obtained from the results of IVR benchmark calculations performed in the IVMR project and presented in (Carénini et al., 2019). Six different codes were used to simulate the same reactor configuration, starting from the steady-state case and then calculating the transient evolution with progressive molten steel incorporation. The heat flux profile at the outer surface of the vessel wall obtained when the maximum heat flux is reached is compared to the heat flux profile obtained in steady-state configuration. All codes show an increase of the heat flux in the transient case. This maximum is located between elevations 1.5 and 2 m in the metal, and corresponds to the initial height of the layer, at a time when the mass of molten steel is still small. The heat flux is multiplied by a factor 1.5-1.9 compared to the simplified steady-state case.

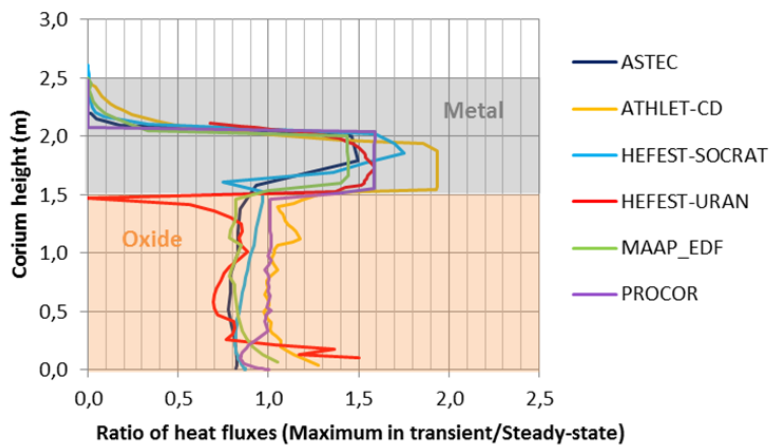

**Figure 1: Multiplying factor obtained between heat fluxes at the outer surface of the vessel calculated based on a steady-state configuration, and the maximum value reached when the transient evolution for the molten steel incorporation in the pool is considered**

Other corium configurations were introduced as in (Rempe et al., 1997) or (Seiler et al., 2007). But the introduction of more configurations, and therefore more parameters in the probabilistic study, made the mathematical analysis more complex and several authors have proposed simpler methods to evaluate the maximum heat flux along the vessel wall. They replaced the complete probabilistic analysis by a selection of conservative assumptions which would provide directly the maximum heat flux as in (Seiler et al., 2007). However, it is not possible to make a conservative assumption for the mass of molten steel, since standard models predict that the maximum heat flux along the metallic layer increases when the height of the metallic layer decreases (so-called focusing effect).

Therefore, the conservative or "bounding case" approach has a limited interest. It is useful to make a quick estimate of the maximum heat flux along the vessel wall, with "reasonably conservative" assumption concerning the minimum thickness of the top metal layer, but it omits deliberately (i.e. assuming a negligible probability for) a whole range of transient situations which could result in higher heat flux, at least temporarily.

Nevertheless, the probabilistic "bounding case" approach introduced initially is still the most widely used approach (Ma et al., 2016). Different approaches may be chosen to evaluate the probabilities of the input parameters and the significance of the probability of the maximum heat flux and minimum wall thickness. In the ROAAM method, the probabilities represent a level of acceptability of an event or physical process with respect to expert judgement or physical sense. Other methods follow an approach similar to PSA and the probabilities are assigned to a frequency of occurrence. It has the advantage of allowing taking into account the frequency of occurrence of some accident scenarios. But, in such a case, the range of frequency covered is continuous and the acceptability criteria are more difficult to define. To illustrate this, in ROAAM approach, an event with a probability lower than 0.01 is considered as "outside the spectrum of reason" (this is a "cut-off" frequency) whereas in a PSA, it would be considered as "possible". Therefore, one has to be very careful when interpreting the results of a probabilistic study of IVR: probabilities may have different meanings.

### 3.1 Important drawbacks of the probabilistic approach

Among the uncertain variables, the mass of metal plays a key role but it is also one of the most difficult to define in terms of distribution. It includes too many sources of uncertainties: it is related to the accident scenario, to the modelling of corium relocation and to the time when steady-state is reached. Moreover, it is not independent of the residual power, which is another uncertain variable. This clearly shows one of the weaknesses of the probabilistic approach: it does not allow taking into account time-dependent variables in a consistent way. Moreover, it assumes that some uncertain variables are independent when they are actually both related to time. As an illustration, it can be seen in Figure 2 that there is clearly a linear correlation between the mass of molten steel from vessel melting and the residual power. Therefore, those variables cannot be considered as independent.

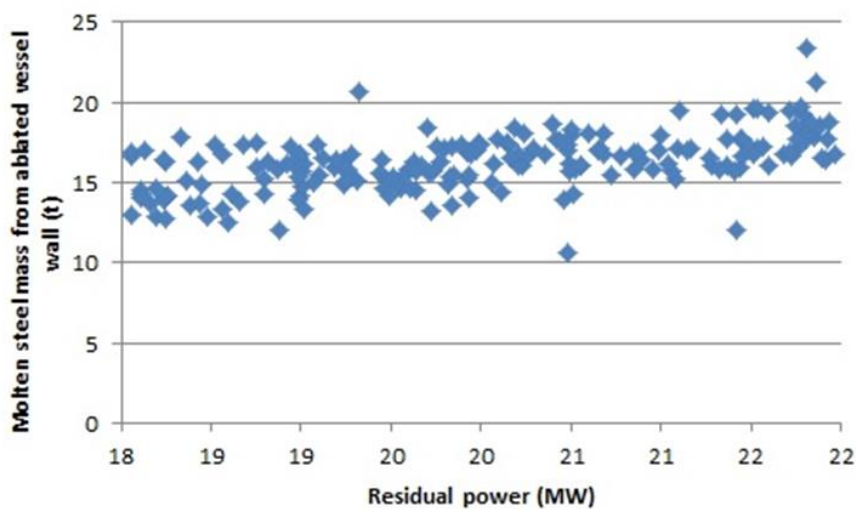

**Figure 2: Variation of the mass of molten steel from the vessel as a function of residual power taking into account transient melting and stratification in the probabilistic evaluation (from ASTEC sensitivity simulations described in IVMR report D2.2.1 (Carénini et al. 2019)**

There is also a possible inconsistency of the probabilistic approach coupled with the "bounding case": the bounding case is "assumed" to be conservative (although it is not) but, if the probability distributions of the uncertain parameters are not defined carefully, it may end up with too much probability for combinations of more favourable parameters (i.e. large mass of steel, low residual power or high emissivity of the metal layer, for instance). In the end, it may lead to overestimate the probability of non-conservative situations.

Finally, another drawback of the probabilistic approach, as it is usually presented, is that it mixes uncertainties related to the scenario and phenomenological uncertainties. This induces confusion in the meaning of the probability of vessel failure. For example, if the result of the probabilistic analysis has a 10% probability of vessel failure: it is a completely different interpretation, if it is found that the vessel would always fail in case of large break which has a probability of 10% or if it is found that the vessel does not fail for any scenario, except with the worse combination of model parameters, which has a probability of 10%.

## 3.2 The issue of mechanical resistance

In the steady-state approach, two criteria for vessel failure were postulated: thermal failure and structural failure. The thermal failure mechanism of the lower head is due to boiling crisis (BC) which occurs when the heat flux through the vessel exceeds the critical heat flux (CHF) at the same location, and it results in a sudden transition of the flow regime from nucleate to film boiling. It has been suggested for an AP600-like reactor (Theofanous 1997) that boiling crisis is a sufficient condition for lower head failure, but also a necessary condition meaning that, unless BC occurs, there can be no failure. This was demonstrated with a detailed mechanical calculation on the most vulnerable point of the vessel, around 90° which means vertical vessel wall, supposing the boundary conditions of Figure 3.

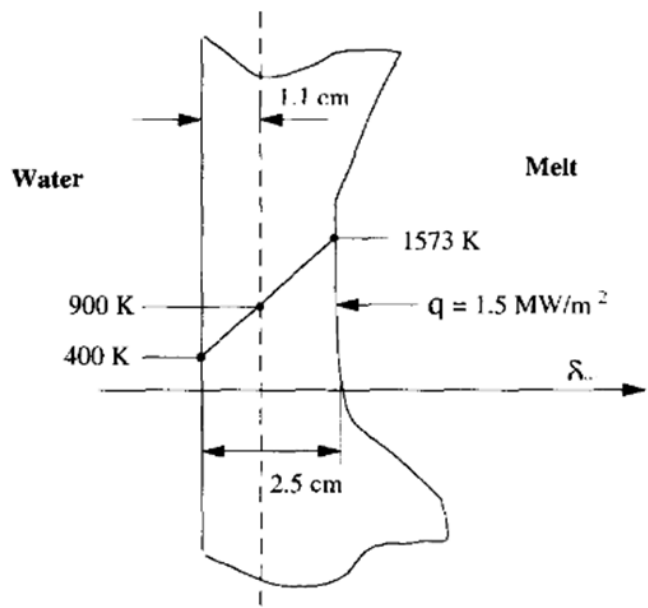

**Figure 3: Illustration of conduction-limited wall thickness, and of the full-strength "cold external shell" region (T < 900 K), under an imposed heat flux of 1.5 MW m-2 (from (Theofanous 1997))**

For the studied cases (Loviisa, AP600), the cold shell was almost 100 times larger than the minimum thickness required to withstand internal overpressure , and the risk of structural failure was soundly rejected.
Analyses, in the IVMR project, of reactors around 1000 MWe power have shown that the thickness of the cold shell may reach values around 1 cm, which is just only 10 times more than the minimum thickness required to withstand static loads of the lower plenum. Conditions like accidental pressurization of the primary circuit or an unfavourable vessel ablation profile reduce this margin further. Thus, for high-power reactors it becomes important to evaluate the margin and check if it is sufficient to deal with the uncertainties on mechanical properties and on the mechanical loading.

## 3.3 Conclusions about existing methodology

The existing methodology appears suitable to demonstrate the efficiency of IVR with ERVC when the safety margin is large (i.e. the residual vessel thickness is almost two orders of magnitude larger than the minimum thickness required to withstand the internal pressure load). Although it is quite comprehensive, this methodology misses an important point: there are transient situations that may cause more damage to the vessel than the supposedly "bounding" case. Moreover, when too many details and parameters are involved, there is also a source of confusion or underestimation in the evaluation of the risk when combining the use of bounding cases and probabilistic assessment. Those drawbacks do not lead to significant errors as long as the safety margin is very large. Reactor calculations performed in the IVMR project support that conclusion. Figure 4 and Figure 5 show the residual thickness of the lower head once the corium is stabilized for a VVER440 and a VVER1000 respectively. In the former case the smallest and most critical residual thickness is about 10 cm, while in the latter it is about 2.5 cm.

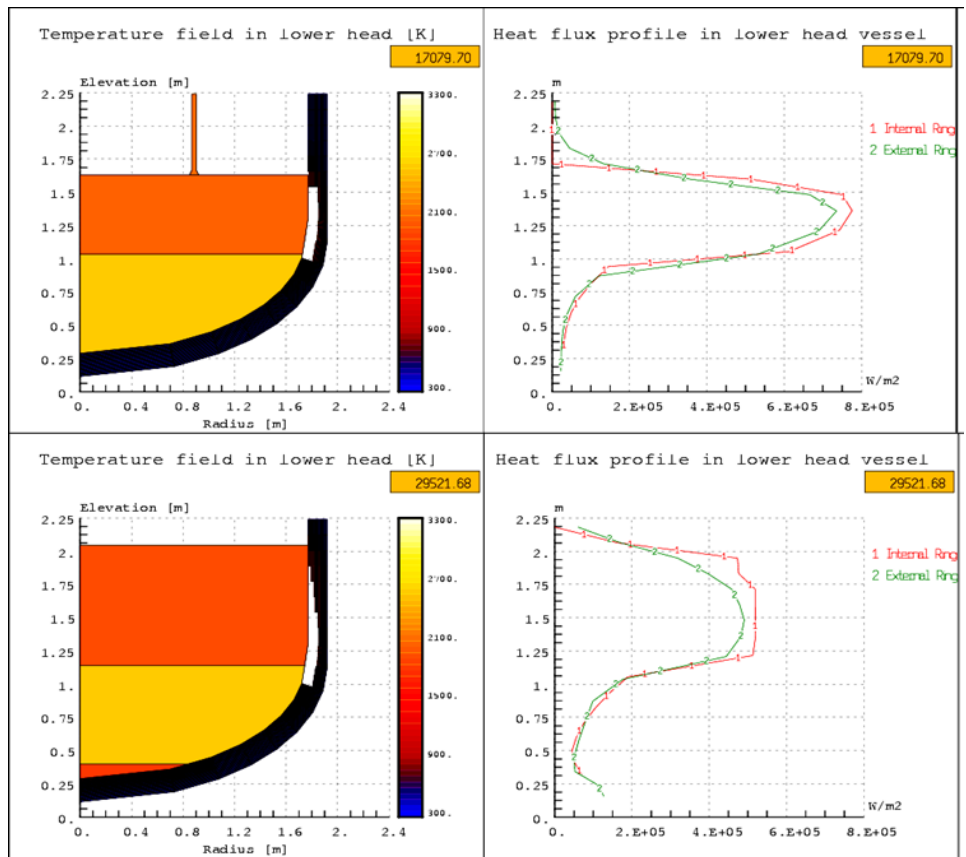

**Figure 4: VV440, LBLOCA, Temperature and HF profile at the time of HF culmination (top) and stabilised state (bottom). Minimum wall thickness is seen to be about 50% of the original RPV thickness (ASTEC simulation by IVS).**

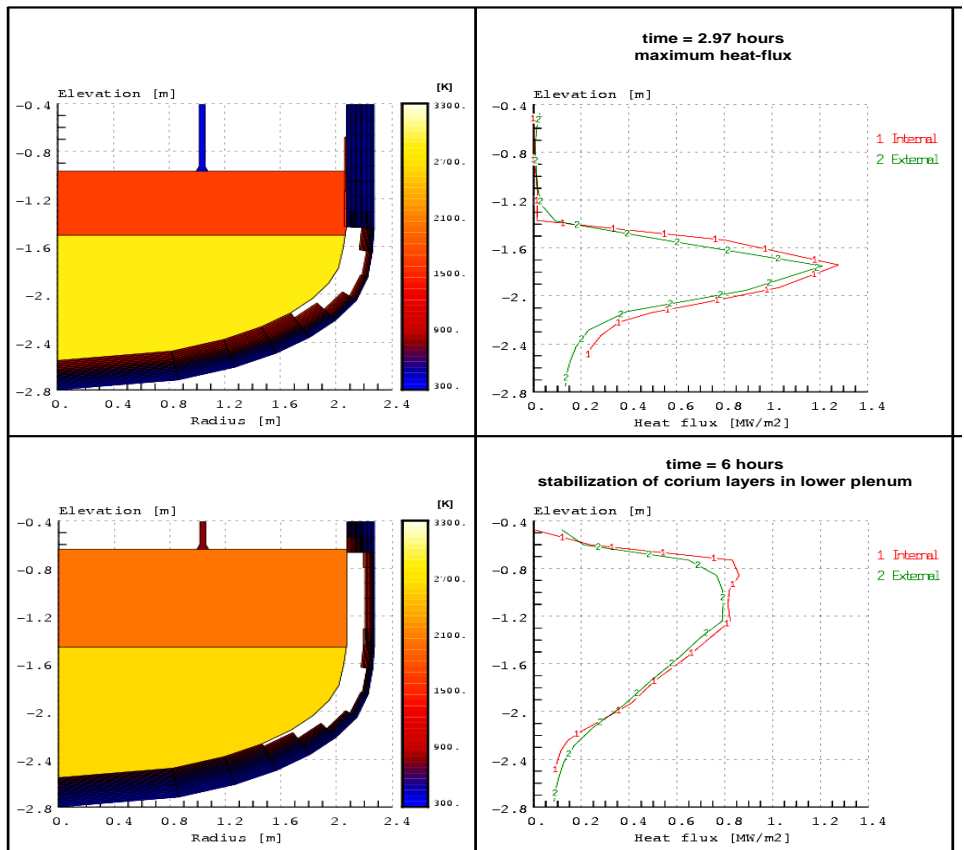

**Figure 5: VVER1000, LBLOCA, temperature and HF profile at the time of HF culmination (top) and stabilised state (bottom). For this latter state - quasi the steady state of the process - the smallest wall thickness caused by the transient process is in a different place than 90°(ASTEC simulations by IVS).**

When the safety margin is reduced, more accuracy is necessary to estimate the margin and all the "details" of the methodology are important. One way of improving both identified drawbacks is to introduce time as a parameter in the evaluation. This can be done in two different ways: either using deterministic (and time-dependent) calculations instead of steady-state cases or, keeping the probabilistic approach on steady-state cases but refining the PDF of uncertain parameters, in order to take into account the fact that most of them are time-dependent and, therefore, they do not have uniform distributions and all combinations of parameters cannot have the same probability. The deterministic way appears more straightforward and more comprehensive and was followed in the IVMR project.

## 4 A key safety parameter: the minimum vessel thickness

As it was shown, the peak heat flux may occur during transient situations, possibly for short time intervals. Nevertheless, this may cause significant ablation of the vessel wall at the location of maximum heat flux. Moreover, the position of maximum heat flux in steady-state may be different from the position of peak transient heat flux. As a result, at any time of the transient, the profile of vessel thickness cannot be deduced only from the knowledge of the heat flux profile in steady-state.
It is clear that the local vessel thickness is a more relevant parameter than the local steady-state heat flux to be used as a criterion of vessel failure because the vessel thickness includes all transient effects since the beginning of the transient, in particular the occurrence of short but very high heat fluxes. It is also a parameter that can be used directly for mechanical calculations over the entire ablated vessel profile.

The residual thickness is a parameter for which it is rather straightforward to estimate a safety margin, as it is usual for other mechanical failure criteria based on thickness/load relationships (such as for the fuel cladding or for the containment). Classically, the safety margin for a structure is defined by a relation such as:

$$\sigma \leq \sigma_{max} = \frac{\sigma_{\text{fail}}}{m}$$

Where σ is the stress and *m* is a safety factor larger than 1. Typically, for standard applications, it can be around 5 when the evaluation of the load is uncertain or if the material properties are not well known. If we have a linear relation between the minimum thickness δ and the stress at failure, then the previous safety margin can be reformulated as:

$$\delta \geq m\ \delta_{\text{fail}}$$

Because of uncertainties on the scenario (possible re-flooding, reliability of the pressure release valves) and on the models (vessel ablation), there are uncertainties in both terms of the thickness/load criterion. Therefore, it is appropriate to choose a factor *m* that is sufficient to cover all possible extreme cases where the residual thickness is reduced by high transient heat fluxes or if the mechanical properties are not well known. In the following parts, evaluations are performed to give orders of magnitude and illustrate the impact of different parameters. The reference values used for calculations are given in the table below.

**Table 1: Example of realistic parameters in an IVR analysis**

| Notation | Description | Reference value | Unit |
|---|---|---|---|
| $\sigma_{cr}$ | Ultimate strength | 600 | [MPa] |
| $R$ | Average radius of the vessel | 2.5 | [m] |
| $T_{\text{fus}}$ | Melting temperature of steel | 1700 | [K] |
| $T_{\text{sat}}$ | Water saturation temperature (outside the vessel) | 400 | [K] |
| $T_{\text{cold}}$ | Maximum temperature of the "cold shell" of the vessel | 700 | [K] |
| $X_{\text{cold}}$ | Relative thickness of the cold shell (w.r.t. total thickness) | 0.23 | [-] |
| $\Delta T_{\text{cold}}$ | Temperature difference across the cold shell | 300 | [K] |
| $\Delta T_{\text{fus}}$ | Temperature difference across the vessel wall | 1300 | [K] |
| $k$ | Vessel steel heat conductivity | 30 | [W/m/K] |

### 4.1 Introducing a new generic safety criterion

To assess the structural integrity in a simple way, we can consider the following 'classical approach'. For the elastic-plastic behaviour, we adopt the failure criterion from (Theofanous et al., 1997) with the assumptions that creep and thermal stresses are neglected:

$$\Delta P_{\max} = \frac{2\sigma_{\text{cr}}\delta_{min}}{R}$$

where $\Delta P_{\max}$ [MPa] is the maximum admissible pressure difference, $\sigma_{\text{cr}}$ [MPa] is the ultimate strength $\delta_{min}$ [m] is the calculated minimum vessel wall thickness, and $R$ [m] is the average radius of the wall.

As it was done before, we may consider that only the "cold part" of the vessel contributes to the mechanical resistance. This can be justified by observing the strong decrease of mechanical properties (Young modules, ultimate strength) when the temperature is above 800K (Willschuetz et al., 2003). In the range 300-800K, the ultimate stress is almost constant and has a minimum value around 600 MPa.

If we consider that the maximum overpressure is $\Delta P_{\max} = 1\ bar$ (i.e. approximately the weight of corium and steel minus buoyancy), we obtain the minimum "cold" thickness:

$$\delta_{cold} = 0.2mm$$

Assuming a linear temperature gradient through the vessel, we can estimate that the calculated vessel thickness leading to rupture is:

$$\delta_{rup}^{min} = \frac{\Delta T_{fus}}{\Delta T_{cold}} \delta_{cold} = 0.87mm$$

Actually, detailed finite element calculations seem to indicate that the criterion based on the plastic rupture of the cold shell is too conservative. The reason is that the "hot part" plays a positive role on the mechanical resistance. This is illustrated in Figure 6 where the limit between safe domain and failure domain evaluated with finite element calculations and estimated with the simple equations introduced above are compared.

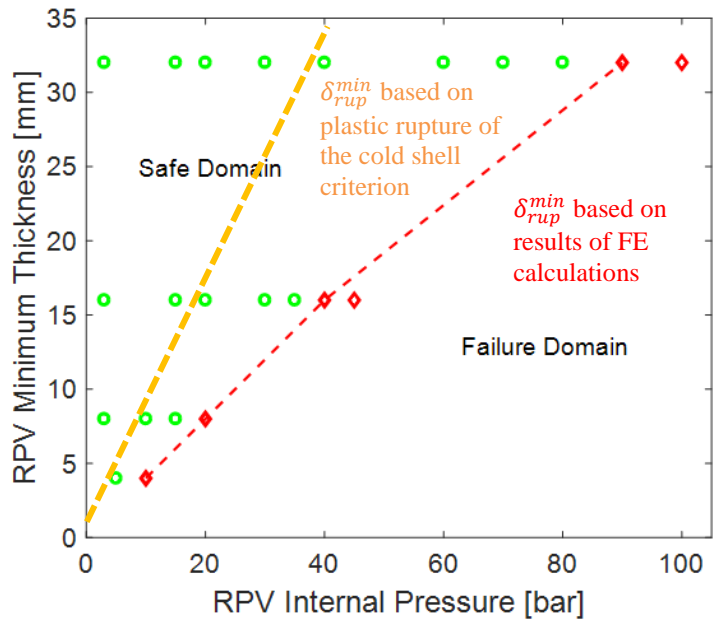

**Figure 6: Comparison of results of finite element calculations (performed by KTH, one point is one calculation) and evaluation performed using the simple mechanical criterion defined in the present report (cf. Eq. 3)**

While a significant decrease of $\delta_{rup}$ values is identified, it is not possible to conclude on a reasonable quantitative value in this report. In addition, uncertainties on mechanical properties (non-homogeneity in the wall, impact of aging in the reactor...) and impact on the value of $\delta_{rup}$ would have to be assessed.

Nevertheless, if we keep the conservative criterion based on the plastic rupture of the cold shell and introduce an additional safety margin $m = 10$, we can re-evalute $\delta_{rup}$ as:

$$\delta_{rup} = 8.7mm$$

Then, a simple safety criterion can be defined as:

$$K_\delta = \delta_{rup}/\delta_{min}.$$

In the example of the VVER440 (Loviisa) where the minimum vessel thickness is always larger than 6 cm we have $K_\delta < 0.145$; this shows that, for an overpressure of 1 bar, there are almost two orders of magnitude for the mechanical safety margin.

## 4.2 Links of steady-state heat flux criterion and thickness criterion

Before comparing the significance of the heat flux criterion and the thickness criterion, it is useful to introduce a new reference value for the heat flux. We define it as the heat flux for which the mechanical failure of the vessel would occur (i.e. when $\delta = \delta_{rup}$):

$$\varphi_{rup} = \frac{k\Delta T_{fus}}{\delta_{rup}}$$

and, for the example of section 3.1

$$\varphi_{rup}(\Delta P_{\max} = 1 bar) \approx 4.5 \, MW/m^2$$

This means that in case of an internal overpressure of, for instance $\Delta P_{\max} = 5 \, bar$, $\varphi_{rup}$ would be reduced to $0.9 \, MW/m^2$. Any higher heat flux would reduce the mechanical safety margin.

At the same time, the maximum heat flux is limited by the capacity of EVRC. In the standard approach, the main "safety" variable is

$$K_\varphi = \varphi_{max}/\varphi_{CHF}$$

What is exactly the meaning of $K_\varphi$ ? If it is above 1, it means vessel failure, without ambiguity. If it is below 1, it represents a "distance" with respect to $\varphi_{CHF}$. But it is not really a safety margin because **there is no classical or standard criterion to determine an acceptable margin for a heat flux**. Therefore, in order to replace that unknown margin factor, the statistical evaluation of this parameter must be done. Then, if $p(K_\varphi > 1) = 0$, it means that there is no risk of vessel failure.

An easy link between both safety criteria can be done by using the ratio $B = \frac{\varphi_{rup}}{\varphi_{CHF}} = \frac{\delta_{CHF}}{\delta_{rup}}$:

1. If $B > 1$, rupture occurs because of excessive heat flux (melt through), irrespective of the internal load
2. If $B < 1$, rupture can occur by mechanical failure before reaching an excessive heat flux

The first case is more realistic, in general, because we assume that the design and accident management measures will allow an effective depressurization. But the second case corresponds, for example, to a scenario where the internal overpressure could reach 30 bar ($\varphi_{rup} \approx 1.5 \, MW/m^2$ (calculated without margin, i.e. $m = 1$), whereas the CHF would be quite efficient, with a value of $2 \, MW/m^2$ or above.

## 5 A revised methodology

In the previous sections, we have seen that, **in order to be general and take into account both risks of mechanical failure and thermal melt-through, it is necessary to consider two safety criteria $K_\varphi$ and $K_\delta$, based on the two parameters $\varphi_{max}$ and $\delta_{min}$**. There are two ways of taking them into account: the "steady-state configuration" approach, which is a fast way to obtain approximate results, and the transient one, which is the most accurate and the best suited way to take into account time-dependent phenomena, in particular in the evaluation of $\delta_{min}$. With both approaches, it is necessary to provide probabilistic results, i.e. distributions of the safety criteria, in order to include uncertainties in the evaluation of the cumulated probability of vessel failure.

**5.1 Characterization of the transient approach**

Considering the points discussed above, it seems obvious that the best way to take into account the uncertainties related to transient processes is to directly calculate vessel ablation in a deterministic way. The vessel thickness is the only variable that is able to keep track of the temporary peaks of heat flux. Another advantage of the deterministic approach is to implicitly take into account the correlations (or dependencies) between several uncertain variables. Among these variables, the most important ones are the mass of steel (which can be calculated as a function of time), the residual power and the internal pressure. This way, the approach is less conservative but more physically grounded as it avoids taking into account non-realistic combinations of variables. The main idea is to use as safety criterion the coupled parameters (minimum vessel thickness / maximum internal load). Of course, this modified methodology is not necessary if the standard criterion on heat fluxes comparison leads to predict failure of the vessel. **The modified methodology is made for a more accurate and rigorous demonstration of the non-occurrence of vessel failure**.

As a first step, the relationship between those two parameters, for which the integrity of the vessel is kept, needs to be tabulated from the results of detailed mechanical calculations, for different types of vessel steel. If such tabulation is not available for the considered steel, a simpler criterion based on the "cold shell" approach can be used (but it is inaccurate). Actually, this task could not be completed in the IVMR project but it should be achieved in the future (for French steel). Then, as a second step, various integral scenarios are calculated, providing curves for the evolution of local vessel thickness as a function of internal load for each angular position along the vessel wall. Finally, this curve is compared to the relationship giving the minimum thickness, checking that it is always reasonably above the minimum thickness, as it is illustrated in Figure 7 for a "virtual" scenario. The different steps of the evaluation are summarized below:

1. Tabulation of minimum vessel thickness $\delta_{min}$
    a. Function of vessel material
    b. Function of internal load: $\delta_{min} = f(P_{int})$
2. Evaluation of internal loads as a function of time
    a. Primary pressure
    b. Corium weight
3. Evaluation of "cumulated" wall ablation as a function of time: $\delta(\theta, t)$ for each angular position $\theta$
    a. Taking into account short peak transient heat flux
    b. Taking into account variation of the angular position of maximum heat flux
4. Checking that $\delta(\theta, t) \gg \delta_{min}$
    a. At any location $\theta$ along the vessel
    b. At any time $t$.

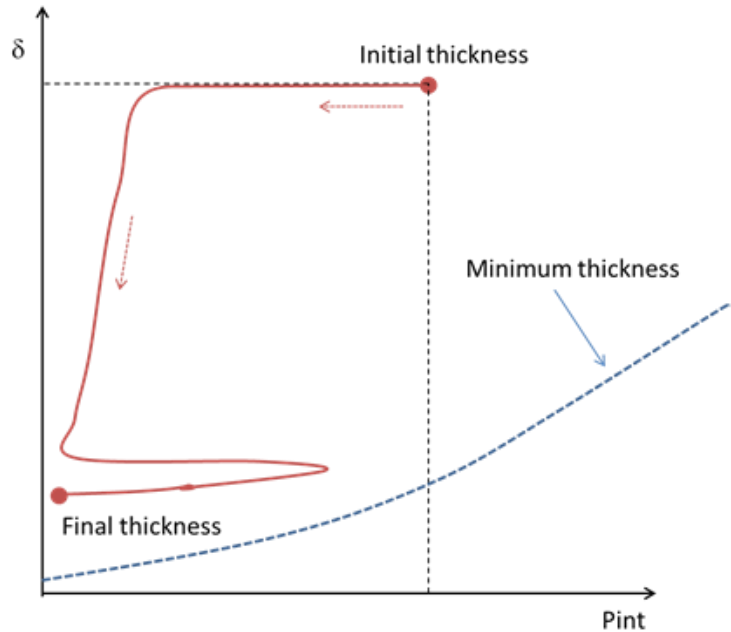

**Figure 7: Example of transient evolutions of local thickness as a function of internal load, compared to the criterion of minimum thickness (fast depressurization followed by a late pressure peak when significant ablation is reached)**

## 6   Conclusions

In the safety evaluation of a reactor design where ERVC is implemented, the first point is to identify if IVR can be achieved with a very high probability of success (i.e. the probability of vessel failure is below a selected threshold) or if there is a non-negligible probability of vessel failure and ERVC is just one mitigation measure among a set of measures, in order to stop the accident progression. In the first case, safety margins must be evaluated very accurately, paying careful attention to the methodology of calculation and to the assumptions made to simplify the problem. In the second case, the result of the evaluation is just an indication about the efficiency of the ERVC to stop the corium within the vessel and the evaluation does not have to be very accurate because the global safety of the reactor depends also on the evaluation of risks induced by ex-vessel phenomena.

For the case where IVR is supposed to be successful in any case, there are two risks to consider: excessive heat flux (i.e. failure by melting) and vessel rupture (mechanical failure). Those two risks can be addressed by comparing the maximum heat flux and the minimum vessel thickness to their respective "reference" values. The higher the reactor power, the more important it becomes to pay a careful attention to the evaluation of the minimum vessel thickness. It is recommended to perform the safety evaluation following a transient approach, because it will provide a better accuracy in the safety margin and it will give a better idea of the most critical situations. The classical "steady-state" configuration approach has many drawbacks, especially when defining a range of variation of coupled and time dependent uncertain parameters in probabilistic evaluation, and may lead to confusing results. It is recommended to use it only as an approximate evaluation when a quick assessment must be done. But it should be completed by transient "best estimate" analyses at least for the most probable scenarios, allowing justification of the assumptions made for the "steady-state" configurations studied.

Now that models are available to describe the transient kinetics of stratification in the oxide/metal pool, and that computation time is not an issue, there is no justification to avoid the transient approach and to keep using "steady-state" configuration approach with confusing or unrealistic assumptions. Moreover, there are more and more indications that SA codes have reached a level of maturity and confidence which allows using them for probabilistic evaluations (Chevalier-Jabet et al. 2013, Mattie et al. 2015, Moiseenko et al. 2013, Wachowiak et al. 2014), and IVR assessment should follow that direction, too.

At the moment, there are still large uncertainties (or missing correlations) for two main phenomena: the heat flux along the top metal layer at the outer surface of the vessel (which is dependent on heat transfers in the thin metal layer but also on the kinetics of metal layer thickness growth) and the minimum vessel thickness before rupture (which might be lower than the value given by the standard "cold shell" model). Improvements are expected soon for the mechanical resistance and the heat transfers in the thin metal layer. It should allow making a more accurate analysis, which is necessary to evaluate high power reactors' safety margins with respect to IVR. For remaining uncertainties, it is recommended to evaluate their impact on IVR evaluation based on the results of sensitivity analysis made with transient calculations. In addition, reactor calculations show that fast sequences (such as LBLOCA) are the ones providing the highest heat flux and vessel ablation but also the highest uncertainties because stratification processes play a more important role. Therefore, it can be concluded that any measure leading to the practical elimination of fast core melt transient would lead to a very large reduction of the probability of vessel failure in case of IVR implementation, by increasing, at the same time, the safety margin and the level of confidence of the evaluation.

## Acknowledgments

The authors wish to acknowledge funding and support provided by the European Project H2020 IVMR No. 662157.

## Nomenclature

| Notation | Description | Unit |
|---|---|---|
| $\delta(t)$ | Local vessel wall thickness | [m] |
| $\delta_{min}$ | Minimum vessel wall thickness | [m] |
| $\delta_{CHF}$ | Vessel wall thickness when $\varphi = \varphi_{CHF}$ | [m] |
| $\delta_{rup}$ | Vessel wall thickness for which mechanical rupture occurs | [m] |
| $\sigma_{cr}$ | Ultimate strength | [MPa] |
| $\sigma_e$ | Effective stress | [MPa] |
| $\Delta P(t)$ | Pressure difference between inside and outside the vessel | [MPa] |
| $Q(t)$ | Residual power per unit mass of $UO_2$ | [W] |
| $R$ | Average radius of the vessel | [m] |
| $T_{fus}$ | Melting temperature of steel | [K] |
| $T_{sat}$ | Water saturation temperature (outside the vessel) | [K] |
| $T_{cold}$ | Maximum temperature of the "cold shell" of the vessel | [K] |
| $X_{cold}$ | Relative thickness of the cold shell (w.r.t. total thickness) | [-] |
| $\Delta T_{cold}$ | Temperature difference across the cold shell | [K] |
| $\varphi(t)$ | Local heat flux along the vessel wall | [MW/m$^2$] |
| $\varphi_{max}$ | Maximum heat flux along the vessel wall | [MW/m$^2$] |
| $\varphi_{CHF}$ | Maximum heat flux that may be extracted by external cooling | [MW/m$^2$] |
| $\varphi_{rup}$ | Heat flux value when $\delta = \delta_{rup}$ | [MW/m$^2$] |
| $M_i$ | Total mass of element "i" | [kg] |
| $H$ | Height of top metal layer | [m] |
| $k$ | Vessel steel heat conductivity | [W/m/K] |

**References**

Carénini L., Fichot F., Bakouta N., Filippov A., Le Tellier R., Viot L., Melnikov I., Pandazis P., Main outcomes from the IVR code benchmark performed in the IVMR project, proceedings of 9th ERMSAR conference (European Review Meeting on Severe Accident Research) Prague, Czech Republic, 18-20 March 2019.

Carénini L., Fichot F., Bakouta N., Filippov A., Le Tellier R., Viot L., Melnikov I., Pandazis P., Code benchmark on IVR: Elaboration, results and main outcomes, D2.2.2 of IVMR project, September 2019.

Carénini L., Fichot F., Bakouta N., Ederli S., Le Tellier R., Pandazis P., Park R.J., Pellegrini M., Peybernes M., Viot L., 2019. Synthesis report on task 2.2: Modelling improvements for Severe Accident computer codes developed during the project and validation works, D2.2.1 of IVMR project, October 2019.

Chevalier-Jabet K., Cousin F., Cantrel L., Séropian C., Source term assessment with ASTEC and associated uncertainty analysis using SUNSET tool. Nucl. Eng. Des. 272, 207-218, 2014.
https://doi.org/10.1016/j.nucengdes.2013.06.042

Esmaili, H., Khatib-Rahbar, M., 2005. Analysis of likelihood of lower head failure and ex-vessel fuel coolant interaction energetics for AP1000. Nucl. Eng. Des. 235, 1583–1605, 2005.
https://doi.org/10.1016/j.nucengdes.2005.02.003

Fichot, F., Bonnet, J., Chaumont, B., IRSN views and perspectives on in-vessel melt retention strategy for severe accident mitigation 1 INTRODUCTION 1–14, 2014

Fichot, F., Carénini, L., Sangiorgi, M., Hermsmeyer, S., Miassoedov, A., Bechta, S., Zdarek, J., Annals of Nuclear Energy Some considerations to improve the methodology to assess In-Vessel Retention strategy for high-power reactors. Ann. Nucl. Energy 119, 36–45. 2018.https://doi.org/10.1016/j.anucene.2018.03.040

Fichot F., Carénini L., Villanueva W., Bechta S., A revised methodology to assess In-Vessel retention strategy for high power reactor, proceedings of ICONE26 conference, London (UK), July 23-26, 2018.

Kymäläinen, O., Tuomisto, H., 2000. CONFIRMING THE IN-VESSEL RETENTION FOR THE LOVIISA PLANT. Proc. Rasplav Semin. 1–8.

Kymäläinen, O., Tuomisto, H., Theofanous, T.G., In-vessel retention of corium at the Loviisa plant, Nucl. Eng. Des. 169, 109–130., 1997. https://doi.org/10.1016/S0029-5493(96)01280-0

Mattie, P.D., Gauntt, R.O., Ross, K., Bixler, N., Osborn, D., Sallaberry, C., Jones, J. and Ghosh, T., State-of-the-Art Reactor Consequence Analyses Project Uncertainty Analysis of the Unmitigated Long-Term Station Blackout of the Peach Bottom Atomic Power Station, U.S. Nuclear Regulatory Commission, NUREG/CR-7155, Washington, DC, 2015.

Moiseenko E. V., Filippov A. S., Ozrin V. D. and Tarasov V. I., BEPU simulation of core melt thermal hydraulics in VVER vessel during the severe accident with SOCRAT/HEFEST and VARIA codes, NURETH-15, Pisa, Italy, May 12-17, 2013.

Rempe, J.L., Knudson, D.L., Allison, C.M., Thinnes, G.L., Atwood, C.L., 1997. Potential for AP600 in-vessel retention through ex-vessel flooding.

Rempe, J.L., Knudson, D.L., Condie, K.G., Suh, K.Y., Cheung, F., Kim, S., 2004. Corium retention for high power reactors by an in-vessel core catcher in combination with External Reactor Vessel Cooling 230, 293–309.
https://doi.org/10.1016/j.nucengdes.2003.11.031

Seiler, J.M., Tourniaire, B., Froment, K., Consequences of material effects on in-vessel retention 237, 1752–1758, 2007. https://doi.org/10.1016/j.nucengdes.2007.03.007

Theofanous, T.G., Liu, C., Additon, S., Angelini, S., Kymäläinen, O., Salmassi, T., In-vessel coolability and retention of a core melt. Nucl. Eng. Des. 169, 1–48, 1997. https://doi.org/10.1016/S0029-5493(97)00009-5

Wachowiak R., Voelsing K., Gabor J., Use of MAAP in Support of Post-Fukushima Applications, EPRI Report 3002001785, 2013.

Wachowiak R., Luxat D., Hanophy J., Kalanich D., Modular Accident Analysis Program (MAAP) - MELCOR Crosswalk Study Phase 1, EPRI Tech report 3002004449, Nov-2014.

Willschuetz, H.-G., Altstadt, E., Sehgal, B.R., Weiss, F., Simulation of creep tests with French or German RPV-steel and investigation of a RPV-support against failure, ANE 30, 1033–1063, 2003. https://doi.org/10.1016/S0306-4549(03)00036-7